\documentclass[10pt,aps,prx,twocolumn,superscriptaddress,nofootinbib]{revtex4-2}
\usepackage[utf8]{inputenc}
\usepackage[T1]{fontenc}
\usepackage{braket}
\usepackage{amsfonts, mathtools}
\usepackage{dcolumn}
\usepackage{booktabs}
\usepackage{siunitx}
\usepackage[breaklinks=true,colorlinks,citecolor=blue,linkcolor=blue,urlcolor=blue]{hyperref}
\usepackage{comment}
\usepackage{orcidlink}
\usepackage[ruled,vlined]{algorithm2e}
\usepackage{xcolor}
\usepackage[normalem]{ulem}
\usepackage{float}
\usepackage{tabularx}
\usepackage{multirow}
\usepackage[T1]{fontenc}
\usepackage[utf8]{inputenc}
\usepackage{placeins} 

\usepackage{listings}
\lstdefinestyle{kipu}{
  basicstyle=\ttfamily\small,
  numbers=left,
  numberstyle=\tiny,
  stepnumber=1,
  numbersep=8pt,
  showstringspaces=false,
  breaklines=true,
  frame=single,
  rulecolor=\color{black!20},
  tabsize=2,
  captionpos=b,
  upquote=true
}
\lstdefinelanguage{json}{
  basicstyle=\ttfamily\small,
  literate=
   *{0}{{{\color{black}0}}}{1}
    {1}{{{\color{black}1}}}{1}
    {2}{{{\color{black}2}}}{1}
    {3}{{{\color{black}3}}}{1}
    {4}{{{\color{black}4}}}{1}
    {5}{{{\color{black}5}}}{1}
    {6}{{{\color{black}6}}}{1}
    {7}{{{\color{black}7}}}{1}
    {8}{{{\color{black}8}}}{1}
    {9}{{{\color{black}9}}}{1}
    {:}{{{\color{black}{:}}}}{1}
    {,}{{{\color{black}{,}}}}{1}
    {"}{{{\color{black}{"}}}}{1},
}

\usepackage{soul}

\begin{document}

\title{Benchmarking Ansatze for Pauli Correlation Encoding in the Maximum Independent Set Problem}

\author{Cian C. Reeves$^{\orcidlink{0009-0006-4258-1845}}$}
\email{cianreeves@kpmg.com}
\affiliation{KPMG US, 2 Manhattan West, New York, NY 10001}

\author{Aaron C. Kemp$^{\orcidlink{0009-0000-6313-0417}}$}
\email{aaronkemp@kpmg.com}
\affiliation{KPMG US, 2 Manhattan West, New York, NY 10001}

\author{Richard Padbury$^{\orcidlink{0009-0003-9772-0521}}$}
\email{rpadbury@kpmg.com}
\affiliation{KPMG US, 2 Manhattan West, New York, NY 10001}

\author{Eva Lia Tarquini$^{\orcidlink{TBD}}$}
\email{etarquini@kpmg.com}
\affiliation{KPMG Switzerland, Badenerstrasse, 172, 8004 Zürich Switzerland}

\author{Christoph Kloeffel$^{\orcidlink{TBD}}$}
\email{ckloeffel@kpmg.com}
\affiliation{KPMG Switzerland, Badenerstrasse, 172, 8004 Zürich Switzerland}

\author{Vedangi Pathak$^{\orcidlink{0000-0001-6335-6539}}$}
\affiliation{IBM T. J. Watson Research Center, 1101 Kitchawan Rd., Yorktown Heights, NY 10598}

\author{Hamed Mohammadbagherpoor$^{\orcidlink{0009-0006-1482-9970}}$}
\affiliation{IBM T. J. Watson Research Center, 1101 Kitchawan Rd., Yorktown Heights, NY 10598}

\author{Vincent Beltrani$^{\orcidlink{0009-0004-9854-4459}}$}
\affiliation{IBM T. J. Watson Research Center, 1101 Kitchawan Rd., Yorktown Heights, NY 10598}


\begin{abstract}
The increasing complexity of 5G and emerging 6G wireless networks has intensified the need for scalable optimization techniques capable of addressing computationally challenging combinatorial problems. While quantum optimization offers a promising alternative, current near-term quantum hardware remains limited by qubit availability, noise, and scalability constraints. Pauli Correlation Encoding (PCE) has emerged as a qubit-efficient optimization framework that addresses these challenges through polynomial variable compression. However, the impact of compression on representability and solution recovery remains poorly understood. In this work, we investigate the hypothesis that increasing compression ratios introduce representability constraints arising from algebraic closure relationships among encoded Pauli observables. Using Maximum Independent Set benchmark problems, we evaluate multiple PCE ansatz designs and compression ratios. We find that ansatz selection is critical to both solution quality and feasibility, with two ansatz families consistently outperforming alternatives and achieving approximation ratios approaching 0.9. Furthermore, increasing circuit repetitions provides limited benefit and can degrade performance. Most notably, reducing compression by increasing the number of available qubits significantly improves raw solution quality, while post-processed performance remains comparatively stable. These findings suggest that representability effects play an important role in compressed PCE encodings and provide new insight into the relationship among compression, ansatz design, decoding, and solution quality in scalable quantum optimization.
\end{abstract}

\maketitle
\section{Introduction}

The evolution of wireless communication systems, particularly the advent of 5G and emerging 6G networks, has introduced unprecedented levels of complexity and stringent performance requirements for the telecommunication industry. These advanced networks demand terabit-per-second data rates, ultra-low latency and massive device connectivity, coupled with improved reliability and energy efficiency\cite{Batool2026QuantumEnhanced, krikidis2025quantumoptimizationwirelesscommunication}. Achieving these ambitious goals requires optimal management of intricate network resources involving critical tasks such as base station placement, beamforming, routing, spectrum allocation and user scheduling\cite{Arani2025Quantum, Bouchmal2025Quantum, Lim2024QuantumAssisted, krikidis2024optimizingconfigurationselectionreconfigurableantenna}. Many of these optimization tasks are inherently combinatorial and computationally challenging nondeterministic polynomial time (NP-hard) problems\cite{Kanza2025Spatially, dinh2022efficienthamiltonianreductionquantum, macaluso2025quantumcomputinglargescalenetwork}. Traditional classical optimization approaches, that includes exhaustive search, become intractable due to their exponential complexity while heuristic and approximation algorithms frequently yield suboptimal solutions by converging to local minima or struggling to meet real-time operational constraints across diverse and dynamic network conditions\cite{Batool2026QuantumEnhanced, Takabayashi2025Optimization, paul2026warmstartquantumapproximateoptimization}. These limitations underscore the pressing need for alternative computational paradigms capable of addressing the increasing demands of modern and future wireless networks.

Quantum optimization methods have emerged as promising alternatives for tackling inherently hard combinatorial optimization problems\cite{Kanza2025Spatially, Bouchmal2025Quantum, krikidis2025quantumoptimizationwirelesscommunication}. Specifically, Quantum Approximate Optimization Algorithm (QAOA) and quantum annealing (QA) are widely used to solve such complex optimization tasks, including related problems in telecommunications applications\cite{krikidis2025quantumoptimizationwirelesscommunication, Batool2026QuantumEnhanced, Takabayashi2025Optimization}. These techniques are frequently integrated into hybrid quantum-classical frameworks that combine quantum processing for complex subproblems with classical computing for overall control and refinement\cite{Kanza2025Spatially, Bouchmal2025Quantum}. A growing body of literature suggests that these quantum and hybrid methodologies hold significant potential for optimizing various aspects of wireless systems, including resource allocation, network planning, routing, scheduling, antenna placement, and beamforming\cite{macaluso2025quantumcomputinglargescalenetwork, dinh2022efficienthamiltonianreductionquantum, Takabayashi2025Optimization, Lim2024QuantumAssisted, Arani2025Quantum, krikidis2024optimizingconfigurationselectionreconfigurableantenna}. These capabilities are directly relevant to critical wireless-network functions such as base station placement, reconfigurable intelligent surface configuration, traffic management, and antenna selection\cite{Arani2025Quantum, Lim2024QuantumAssisted, krikidis2024optimizingconfigurationselectionreconfigurableantenna, vandelli2025constraintpreservingquantumalgorithmmultifrequency}.

Despite the promise of quantum optimization, its practical application to large-scale wireless systems faces significant hurdles primarily due to the current limitations of quantum hardware. Today's near-term quantum devices are characterized by a restricted number of qubits and susceptibility to noise and decoherence\cite{bharti2022noisyintermediate,Chen2023thecomplexityofnisq}. This challenge is particularly pronounced in algorithms such as QAOA, where increasing problem sizes typically require deeper quantum circuits, leading to greater error accumulation and reduced solution reliability\cite{Bouchmal2025Quantum, krikidis2024optimizingconfigurationselectionreconfigurableantenna, paul2026warmstartquantumapproximateoptimization, macaluso2025quantumcomputinglargescalenetwork, Batool2026QuantumEnhanced}. Consequently, scaling quantum optimization methods to address real-world problems, which often involve millions of variables and complex network topologies, remains a profound challenge\cite{Kanza2025Spatially}. In particular, the process of mapping problem variables to physical qubits from multi-constrained objectives often requires a higher number of physical qubits per variable, which further stresses limited hardware resources as problems increase in complexity\cite{Takabayashi2025Optimization, dinh2022efficienthamiltonianreductionquantum}. Therefore, bridging the gap between the theoretical potential of quantum optimization and its practical, scalable implementation for next-generation wireless systems necessitates sustained innovation in both quantum hardware and algorithmic design.

In response to the computational intractability of classical methods for certain problems and the constraints imposed by current near-term hardware, Pauli Correlation Encoding (PCE) emerged as a qubit-efficient variational framework \cite{Original_Sciorilli_2025, PadinMartinez2026Progressive, Galda_mRNA_2026paulicorrelationencodingmrna}. As demonstrated in the original work of Sciorilli et al., PCE directly addresses the challenge of mapping many classical binary variables onto limited quantum resources by encoding them into the expectation values of Pauli observables.\cite{Original_Sciorilli_2025} Rather than dedicating a qubit to each variable, this innovative approach enables polynomial compression allowing $m$ binary variables to be represented using $n$ qubits, where $n$ scales polynomially with $m$ (e.g., $O(m^{1/k})$ for a tunable compression order $k \geq 2$)\cite{Original_Sciorilli_2025, alonso2026benchmarkpaulicorrelationencoding, Galda_mRNA_2026paulicorrelationencodingmrna, PadinMartinez2026Progressive, bosco2026efficientlysimulablepaulicorrelation}. Such enhanced qubit utilization and variable compression are crucial for tackling problems where traditional one-hot encoding would quickly exceed the capabilities of current quantum devices\cite{nguyen2026scalingquantumoptimizationunit}. Furthermore, PCE has been analytically demonstrated to mitigate the problematic barren plateau phenomenon, offering super-polynomial resilience in training landscapes, which is a significant barrier for variational quantum algorithms\cite{Original_Sciorilli_2025, soloviev2026quantumlearningclassicalcorrelations, carmo2025warmstartingpcetravelingsalesman}. As such, PCE has been explored for mRNA secondary structure prediction (enabling the study of sequences involving hundreds of variables with a few tens of qubits)\cite{Galda_mRNA_2026paulicorrelationencodingmrna}, portfolio optimization problems with over 250 assets\cite{Soloviev2026LargeScale}, energy systems for unit commitment problems (compressing hundreds of binary variables onto a significantly smaller qubit count)\cite{nguyen2026scalingquantumoptimizationunit} and for budget-constrained optimization in areas like min-cut\cite{PadinMartinez2026Progressive} and the Traveling Salesman Problem \cite{carmo2025warmstartingpcetravelingsalesman}. Additionally, PCE has been benchmarked across general combinatorial optimization problems including Max-Cut and Bin Packing, showcasing its broad applicability\cite{alonso2026benchmarkpaulicorrelationencoding}.

Despite these advances, several challenges remain, including the difficulty of translating continuous expectation values into feasible binary solutions, particularly for problems with dense constraints, as well as the substantial measurement overhead associated with estimating large numbers of Pauli expectation values\cite{PadinMartinez2026Progressive, bosco2026efficientlysimulablepaulicorrelation, Galda_mRNA_2026paulicorrelationencodingmrna}. Moreover, the selection and tuning of hyperparameters like the binarization parameter alpha remain critical and problem-specific, impacting solution quality and feasibility\cite{PadinMartinez2026Progressive, alonso2026benchmarkpaulicorrelationencoding}. The existence of classically simulable PCE variants also necessitates careful benchmarking to delineate genuine quantum advantage\cite{bosco2026efficientlysimulablepaulicorrelation}. Furthermore, important questions remain regarding the fundamental representational properties of compressed Pauli encodings. Recent work has begun to formalize representability constraints in such encodings, showing that algebraic dependencies among encoded observables can induce closure relations that restrict the set of physically realizable variable assignments. As a result, not all binary configurations may be accessible within a given encoding\cite{bosco2026efficientlysimulablepaulicorrelation}. At the same time, the literature has primarily focused on improving decoding and constraint-satisfaction through methods such as local search, Progressive Binarization, and Problem-Aware Guided Decoding (PAGD), with comparatively limited investigation into how representability constraints evolve as compression increases\cite{bosco2026efficientlysimulablepaulicorrelation, Galda_mRNA_2026paulicorrelationencodingmrna}.

Nonetheless, PCE's ability to efficiently encode complex optimization problems onto limited qubit architectures, coupled with its demonstrated resilience to barren plateaus, positions it as a promising framework for developing scalable quantum optimization methods for challenging domains such as next-generation wireless systems. Therefore, in this study, we investigate the maximum independent set problem, directly relevant for telecommunications, to systematically evaluate PCE across multiple ansatz designs and compression ratios. We examine whether increasing compression amplifies representability constraints and explore how these constraints influence decoding behavior, solution recovery, and overall solution quality.

\section{Methods}
In this section we will discuss the PCE approach in more detail, introduce the ansatze we test, describe the problem we are studying and provide details on the numerical simulations carried out in this work.
\subsection{Pauli Correlation Encoding}
As mentioned in the introduction, Pauli Correlation encoding (PCE) is a qubit efficient scheme for encoding binary variables onto a quantum computer.  The core idea of PCE is to encode binary variables ($x_i$) into the signs of Pauli string expectation values ($\langle \Pi_i\rangle$).  
\begin{equation}\label{eq:sgn_encoding}
    x_i = \textrm{sgn}(\langle \Pi_i \rangle ) \quad\forall i \in \{1,\dots,m\}
\end{equation}
where $m$ is the number of binary variables to be encoded.  In principle $\langle \Pi_i\rangle$ can be any combination of Pauli operators\cite{Original_Sciorilli_2025,bosco2026efficientlysimulablepaulicorrelation}, however this can come with significant measurement overhead. By only allowing each string to contain one type of Pauli operator, i.e only $X$'s, only $Y$'s or only $Z$'s, all expectation values can be found with only three measurement settings.  This measurement efficient approach is typically applied\cite{Original_Sciorilli_2025,PadinMartinez2026Progressive,soloviev2026quantumlearningclassicalcorrelations,Soloviev2026LargeScale,Galda_mRNA_2026paulicorrelationencodingmrna,alonso2026benchmarkpaulicorrelationencoding} and is used in this work. The order of the Pauli string refers to the number of non-identity operators that appear in the string, e.g. $ZZII$ is an order two Pauli string and $IXIXX$ is an order three Pauli string.  The order of the Pauli strings determines how many binary variables can be encoded.  For order $k$ strings on $N$ qubits there can be $N_\textrm{total strings} = 3\left(N\atop k\right)$ binary variables encoded, coming from the number of possible placements of $k$ non-identity Paulis among $N-k$ identity operators, the factor of 3 comes from the fact that we can have $X, Y$ and $Z$ strings.  $k$ is typically kept at 2 or 3 which are found to provide good results\cite{Original_Sciorilli_2025, alonso2026benchmarkpaulicorrelationencoding, Galda_mRNA_2026paulicorrelationencodingmrna, PadinMartinez2026Progressive}.

In practice the sgn encoding used in equation \eqref{eq:sgn_encoding} is not good for optimization since the sgn function is not smooth.  For this reason sgn is replaced by a smooth function that has the same asymptotic behavior, such as tanh or the sigmoid function.  With this change the cost function proposed in the original PCE paper\cite{Original_Sciorilli_2025} is
\begin{equation}\label{eq:pce_cost_function}
    \begin{split}
        \mathcal{L} = \sum_{(i,j) \in E} w_{ij}\tanh(\alpha \langle& \Pi_i \rangle ) \tanh(\alpha \langle \Pi_j \rangle ) + \\&\beta \nu \left[\frac1m \sum_{i\in V} \tanh(\alpha \langle\Pi_i\rangle)^2\right]^2.
    \end{split}
\end{equation}
Here $E$, $V$, $w_{ij}$ denote the set of edges, nodes and weights of the corresponding quadratic unconstrained binary optimization (QUBO) problem.  The first term represents a relaxation of the original QUBO problem with $x_i \rightarrow \tanh(\alpha\langle \Pi_i\rangle)$.  The second term is introduced as a regularization term that keeps the correlators close to zero.\cite{Original_Sciorilli_2025}  This improves the solvers capability and will be discussed more in section \ref{sec:discussion}.  $\alpha$ is a hyperparameter introduced to allow $\tanh(\alpha \langle \Pi_i \rangle)$ remain in a non-linear regime even when the correlator is small, which is helpful for the optimizer, as well as to help binarize the variables, i.e. to help ensure $x_i \rightarrow \{0,1\}$.  $\nu$ estimates the cut value of the QUBO problem so that the regularization term is on the order of the first term in the expression.  Following \cite{Original_Sciorilli_2025} the Poljak-Turzik lower bound is used $\nu = w(G)/2 + w(T_\mathrm{min})/4$, where $w(G)$ and $w(T_\mathrm{min})$ are the weights of the graph and it's minimum spanning tree respectively.  In the results shown here we fix $\beta = 0.5$ and $\alpha = N_\mathrm{qubits}$, these were found to give good results in\cite{Original_Sciorilli_2025}.  

The PCE proceeds by choosing an ansatz in the form of a parameterized quantum circuit.  Then one iteratively runs the circuit and measures the Pauli strings chosen for the encoding.  These are fed into equation \eqref{eq:pce_cost_function} and a classical optimizer updates the ansatz parameters until convergence is achieved. Typically, an efficient classical post-processing is performed on the raw PCE bit strings as this often leads to large improvement.  For a more in-depth discussion of the PCE and variations on it see\cite{Original_Sciorilli_2025, PadinMartinez2026Progressive,alonso2026benchmarkpaulicorrelationencoding}.

\subsection{Ansatze}\label{sec:ansatze}
\begin{figure}
    \centering
    \includegraphics[width=\linewidth]{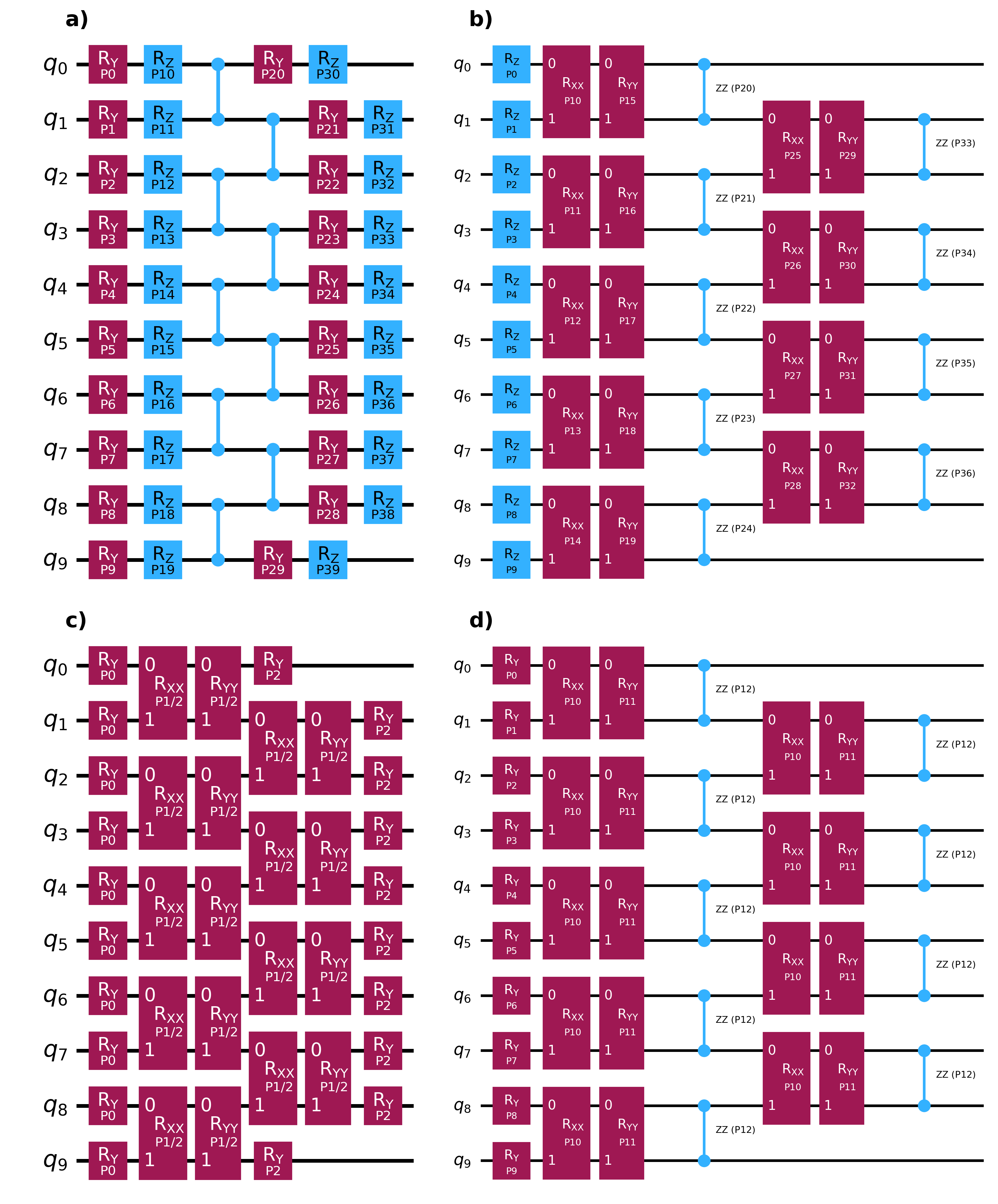}
    \caption{Single repetition of each ansatz tested in this work. The last set of single qubit rotations in Fig. \ref{fig:tested_ansatz} a) and c) are not repeated for every repetition but just placed at the very end of the circuit independent of the number of repetitions.}
    \label{fig:tested_ansatz}
\end{figure}

Each ansatz we test uses parameterized single qubit rotation and two qubit entangling gates laid out in a typical brickwork structure.  A single repetition of each of the tested ansatz is shown in Fig. \ref{fig:tested_ansatz}.  We note that for the ansatze in Fig. \ref{fig:tested_ansatz} a) and c) the final rotation layer is not applied in every repetition but only added to the end of the entire circuit.   Ansatz A [Fig. \ref{fig:tested_ansatz} a)] has similar structure to the ansatz applied in early PCE studies\cite{soloviev2026quantumlearningclassicalcorrelations,Soloviev2026LargeScale}. Ansatz C [Fig. \ref{fig:tested_ansatz} c)] has been applied to the problem of secondary structure prediction in mRNA\cite{Galda_mRNA_2026paulicorrelationencodingmrna}. Notably this ansatz only has one parameter per layer making it very efficient to optimize at the cost of it being less expressive than other ansatze with more parameters.  Ansatz B and D [Fig. \ref{fig:tested_ansatz} b) and d)] are inspired by the Heisenberg Hamiltonian. We believe the direct coupling of $XX$, $YY$ and $ZZ$ Pauli operators on adjacent sites may help in generating correlations between operators that make up the Pauli correlation encoding. Ansatz B and D differ in two ways, firstly the single qubit rotations are $Rz$ and $Ry$ respectively, secondly ansatz D uses only one parameter per entangling gate type in a given layer, significantly reducing the parameter count compared to ansatz B which has a unique parameter for each gate.  Ansatz D gains back expressivity over ansatz C by having a parameter for each single qubit rotation gate as well as by having a different parameter for each type of entangling gate in a given layer ($XX$ vs $YY$ vs $ZZ$) instead of one parameter for the whole layer.  These changes allow ansatz D to scale well and to maintain expressivity without choking the optimizer with too many parameters. As we will see in section \ref{sec:ansatz_testing} both problems can lead to poor results from PCE in the other ansatze we have tested.  Table \ref{tab:parameter_scaling} shows a comparision between the number of parameters each ansatz has as well as the depth after transpilation to IBM quantum hardware (Marrakesh) with optimization level 3.
\begin{table*}
    \centering
    \begin{tabular}{||c||c|c|c|c||}\hline
        &Ansatz A & Ansatz B & Ansatz C & Ansatz D\\
        \hline
        Parameter count &$2N_q(N_\mathrm{reps}+1)$ &$(4N_q-3)N_\mathrm{reps}$&$2N_\mathrm{reps} + 1$&$(N_q+3)N_\mathrm{reps}$ \\
        \hline
        Transpiled depth &$\sim$ 12 &$\sim$ 47&$\sim$ 42&$\sim$46\\
        \hline
    \end{tabular}
    \caption{Parameter count for each ansatz as a function of the number of ansatz repetitions and number of qubits and circuit depth after transpilation for IBM quantum hardware with optimization level 3 for each ansatz shown in Fig. \ref{fig:tested_ansatz}}
    \label{tab:parameter_scaling}
\end{table*}

\subsection{Test Systems}\label{sec:test_systems}
To test the different ansatze described in Section \ref{sec:ansatze} we apply PCE to the maximum independent set (MIS) problem.  The MIS problem is a constrained binary optimization problem where the goal is to find the maximum set of disjoint nodes in a graph.  It has applications in resource allocation, job scheduling as well as telecommunications and networking, where optimal solutions must be found while accounting for conflicts and interference between elements.  The MIS problem can be mapped to a QUBO problem by reformulating the independent set constraint as a term in the cost function that penalizes sets containing connected nodes.  This softens the constraint and local minima of the optimization problem can correspond to invalid MIS solutions, i.e. they correspond to non-independent sets.  This means we not only look the MIS size as a measure of the ansatze capability but also whether or not they produce feasible MIS solutions.  

We perform simulations for three different graph sizes, with sizes corresponding with the number of available order two Pauli strings for $N_q=10$ (135 nodes), $N_q=15$ (315 nodes) and  $N_q=20$ (570 nodes).  The specific mapping used here, to take the MIS to a QUBO problem (max-cut), introduces one additional graph node to the problem and so the original MIS problems each have one less node than the values above.  The nine MIS graphs are shown in Fig. \ref{fig:tested_graphs} in appendix \ref{sec:graph_instances}
\subsection{Numerical Details}\label{sec:numerical_details}
Each graph we study is randomly generated by selecting points on a grid separated by some minimum distance, and introducing a parameter that determines the distance between nodes that connects them with an edge.  The nine graphs we tested in this paper are shown in Fig. \ref{fig:tested_graphs} in appendix \ref{sec:graph_instances}.   We use the CPLEX solver to get the globally optimal MIS size for each graph to benchmark our ansatze against.  For each ansatz we perform simulations with one, two and three repetitions of the blocks shown in Fig. \ref{fig:tested_ansatz}.  We perform classical simulations of PCE, computing the expectation values that enter  equation \eqref{eq:pce_cost_function} using IBM Qiskit's Aer simulator with the matrix product state method.  The classical optimization was performed with the COBYLA optimizer with the maximum iterations (max\_iter) set to max(125, \# optimization parameters), the initial trust region size(rhobeg) set to 0.3 and the final size of the trust region (tol) set to $10^{-3}$.  Using numpy's uniform random number generator, with upper and lower bounds of $\pm 0.4$, we randomly select 5 sets of initial circuit parameters for each system and ansatz studied.  We found that having a narrower random uniform range helped with convergence of the optimizer.

After decoding the Pauli-correlation expectation values into a binary solution using a sign-based rule, we applied a classical one-bit-flip local search as a post-processing step\cite{Original_Sciorilli_2025}. Each binary variable was sequentially inverted and the objective function was re-evaluated. A bit flip was accepted whenever it improved the objective value; otherwise, it was reverted. The procedure terminated after a complete pass with no further improving moves, yielding a locally optimized solution.

\section{Results}
\subsection{Ansatz Testing}\label{sec:ansatz_testing}
\begin{figure*}
    \centering
    \includegraphics[width=\linewidth]{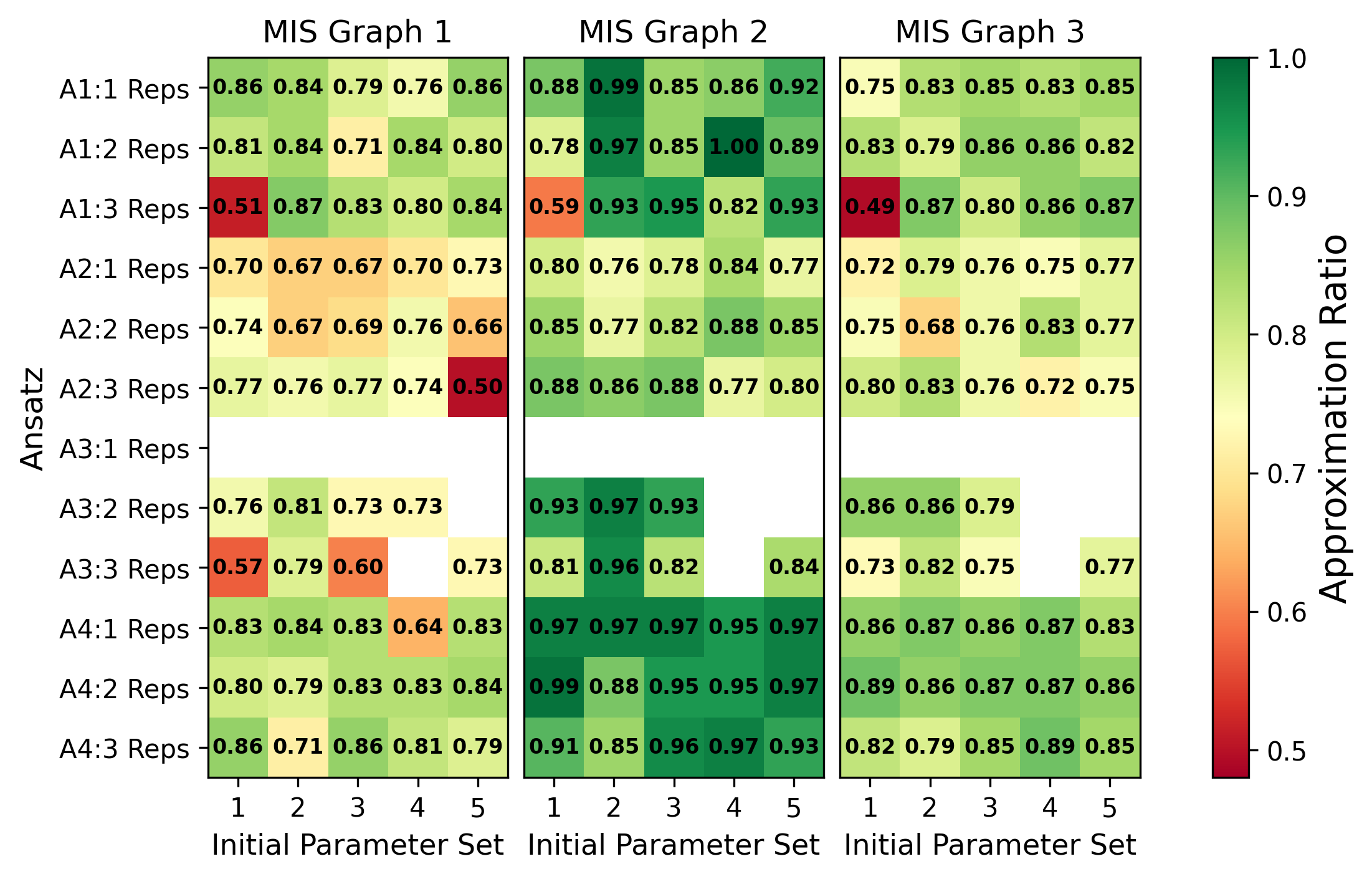}
    \caption{Approximation ratio produced by PCE, after post-processing, for each ansatz in Fig. \ref{fig:tested_ansatz} for 10 qubits corresponding to a 134 node MIS graph[specific instances are shown in Fig. \ref{fig:tested_graphs} a)-c)].  Ratios are with respect to the exact optimum computed with CPLEX.  Blank spaces indicate the solution was not a valid independent set.  The $y$-axis corresponds to the different ansatze and different number of repetitions, the $x$-axis corresponds to different initial parameters used to start the classical optimization.}
    \label{fig:10_qubits}
\end{figure*}

\begin{figure*}
    \centering
    \includegraphics[width=\linewidth]{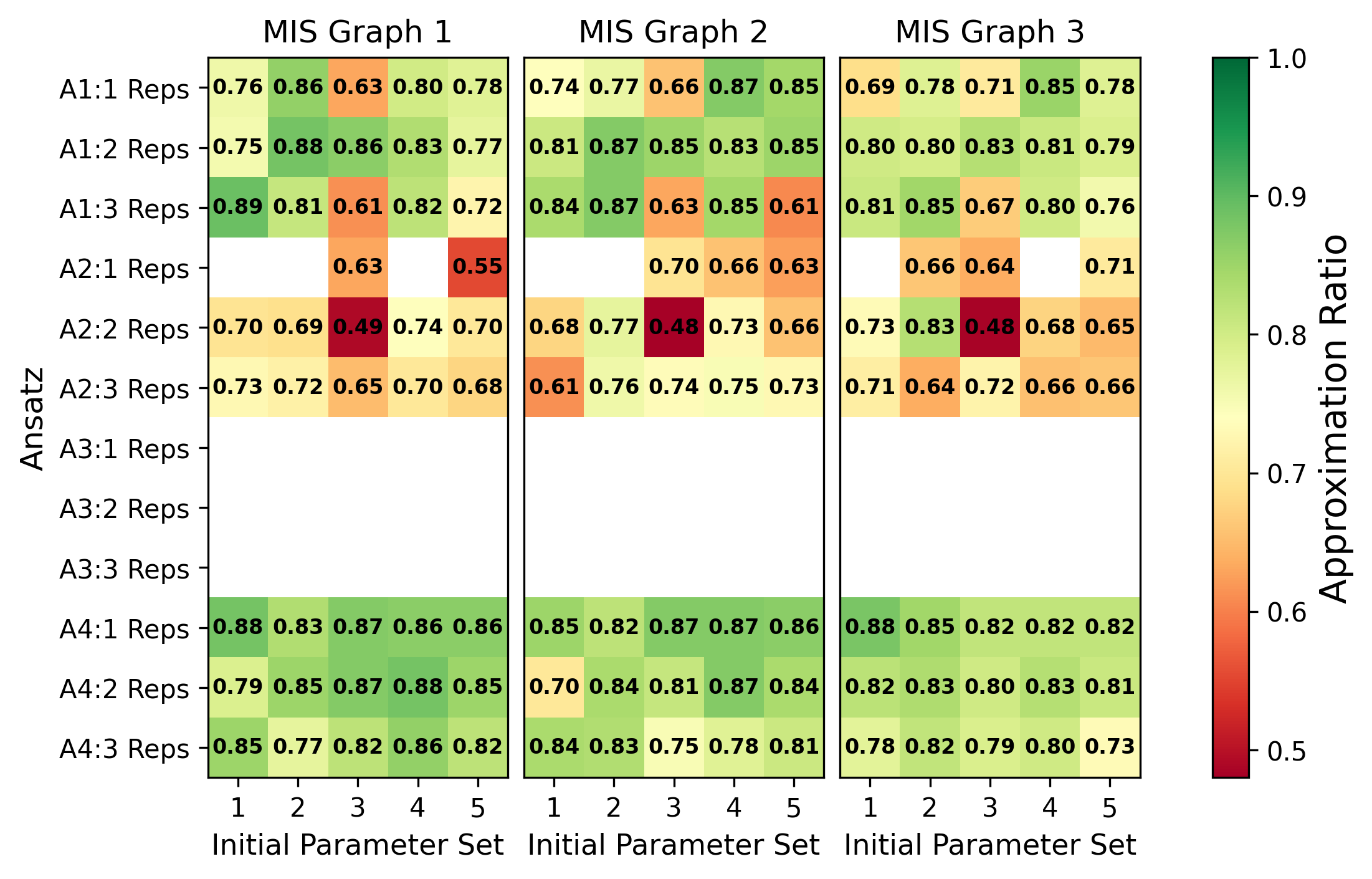}
    \caption{Approximation ratio produced by PCE, after post-processing, for each ansatz in Fig. \ref{fig:tested_ansatz} for 15 qubits corresponding to a 314 node MIS graph[specific instances are shown in Fig. \ref{fig:tested_graphs} S1 d)-f)].  Ratios are with respect to the exact optimum computed with CPLEX.  Blank spaces indicate the solution was not a valid independent set.  The $y$-axis corresponds to the different ansatze and different number of repetitions, the $x$-axis corresponds to different initial parameters used to start the classical optimization.}
    \label{fig:15_qubits}
\end{figure*}

\begin{figure*}
    \centering
    \includegraphics[width=\linewidth]{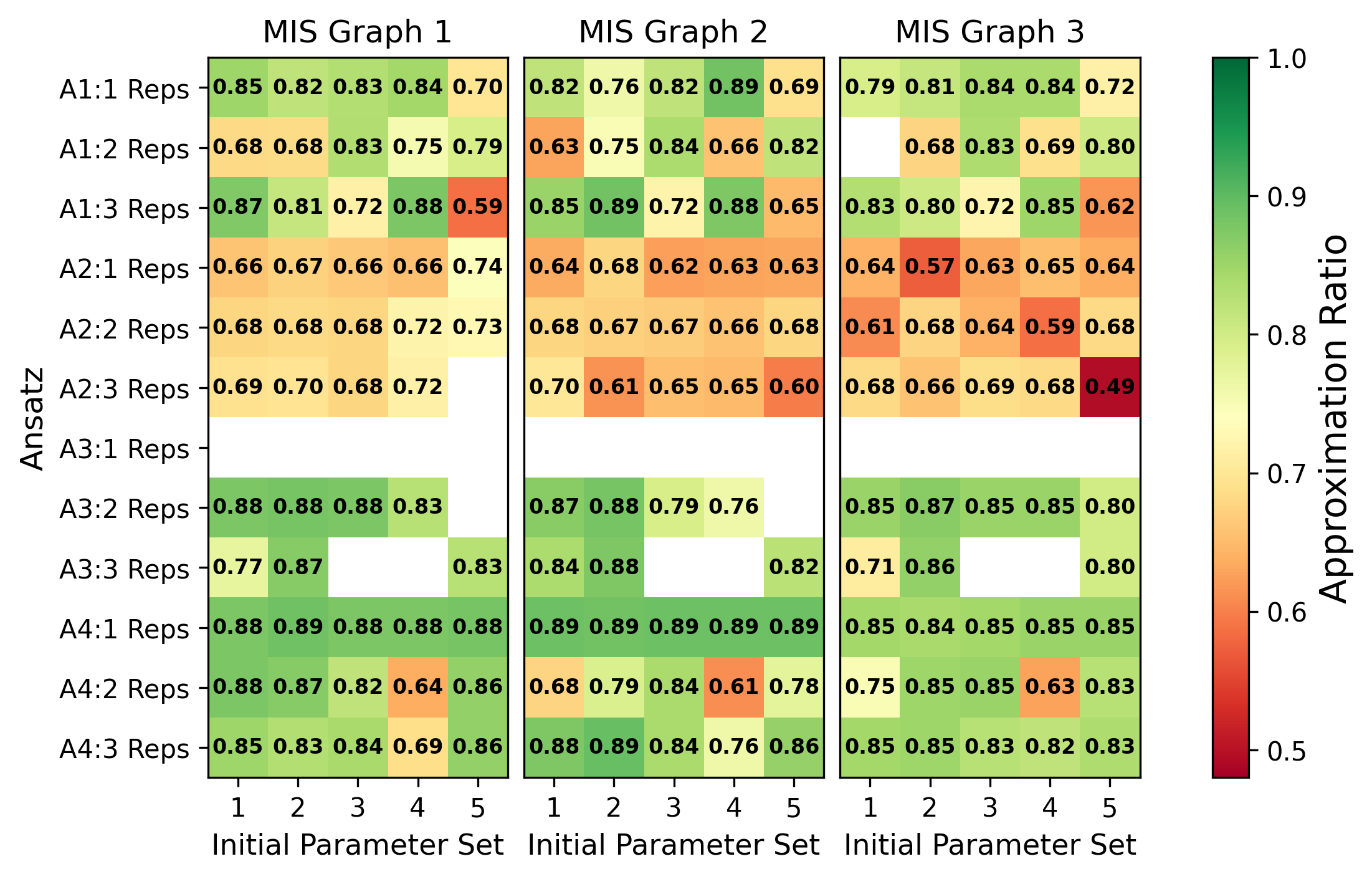}
    \caption{Approximation ratio produced by PCE, after post-processing, for each ansatz in Fig. \ref{fig:tested_ansatz} for 20 qubits corresponding to a 569 node MIS graph[specific instances are shown in Fig. \ref{fig:tested_graphs} g)-i)].  Ratios are with respect to the exact optimum computed with CPLEX.  Blank spaces indicate the solution was not a valid independent set.  The $y$-axis corresponds to the different ansatze and different number of repetitions, the $x$-axis corresponds to different initial parameters used to start the classical optimization.}
    \label{fig:20_qubits}
\end{figure*}

In this section we present results comparing the ansatze shown in Fig. \ref{fig:tested_ansatz} in different instances of maximum independent set problem(MIS) (shown in Fig. \ref{fig:tested_graphs} a)-c)).  Details on the MIS problem and the instances tested are given in sections \ref{sec:test_systems} and \ref{sec:numerical_details}.  Figs. \ref{fig:10_qubits},\ref{fig:15_qubits} and \ref{fig:20_qubits} show the approximation ratio with respect to the exact optimum computed with CPLEX across different ansatz, ansatz repetitions, MIS graph instances and initial parameter settings for 10, 15 and 20 qubits respectively.  The results shown are after classical post-processing, details of the post processing procedure are given in Section \ref{sec:numerical_details}.  White squares indicate that the final solution(after post-processing) was not a feasible solution, i.e. not a valid MIS.  The calculations presented here were done on an ideal matrix product state simulator.
First we look at Fig. \ref{fig:10_qubits}, corresponding to three 134 node MIS problems[Fig. \ref{fig:tested_graphs} a)-c)],  encoded onto 10 qubits using the PCE. We see that ansatze A and D have the best performance across the tests.  Ansatz A produces approximation ratios up to 0.87, 1.00 and 0.87 for each of the three graphs respectively while ansatz D performs similarly, with best ratios of 0.86, 0.99 and 0.89 for the same 3 graphs.  In comparison, ansatze B and C perform noticeably worse.  Ansatz B produces approximation ratios consistently 5-15\% lower than those produced by ansatze A and D, with it's best results being 0.76, 0.88 and 0.83 for each MIS instance respectively.  For ansatz C although the best results across the 3 MIS instances, 0.81, 0.97 and 0.86, are on par with the best results of ansatze A(0.87,1.00,0.87) and D(0.86,0.99,0.89), over half of the solutions it produces are infeasible. Adding more repetitions of ansatz C alleviates this problem but does not eliminate it entirely.  For ansatz A and D increasing the number of repetitions has a slight diminishing effect on the results while for ansatz B a slight improvement is seen.

Moving to the results in Fig. \ref{fig:15_qubits}, corresponding to three 314 node MIS instances encoded onto 15 qubits and shown in Fig. \ref{fig:tested_graphs} d)-f)  we see a noticeable change in the results.  The most striking observation is that ansatz C now produces no feasible solutions even for the maximum tested number of repetitions.  Ansatz B now produces infeasible solutions as well when a single repetition is used, however for two and three repetitions feasible solutions are again found. Of the three ansatz that produce results ansatz B performs the worst, comparing it to ansatz A and D we see that the latter are consistently 10-15\% higher than the results produced by ansatz B.  Even the best results produced by ansatz B, 0.73, 0.77 and 0.83, are significantly lower than the best produced by ansatz A, 0.89, 0.87 and 0.85, and ansatz D, 0.88,0.87 and 0.88.  The best results of ansatze A and D across the three graphs are again comparable in quality, but a notable difference is that ansatz D gives more consistently good results.   For ansatz D, within each graph, the solutions fall within 0.11, 0.17 and 0.15 of one another, while for ansatz A the range is 0.28, 0.24 and 0.18.
In these cases, increasing the number of repetitions has a different effect depending on the ansatz.  The most notable is for ansatz B where increasing the number of repetitions from one to two and three removes the previous infeasible MIS solutions.  Additionally, although there is not significant improvement of the typical values by increasing the repetitions of ansatz B there is an improvement in consistency.  This is seen for example in MIS graph 2 where the solutions for two repetitions can differ by as much as 0.29 whereas this drops to only 0.15 difference across solutions when three repetitions are used.  This is similar for the other two graphs tested at this size.  For ansatz A we see a similar story, namely moving from one to two repetitions of the ansatz we see the best values remain similar but the worst solutions improve by 15-35\%, however this trend does not continue when a third repetition of ansatz A is introduced.  Interestingly, for ansatz D increasing the number of repetitions immediately leads to worse results which continues for ansatz D as we increase from two to three repetitions, although the change between the three repetitions is not significant .

Fig. \ref{fig:20_qubits} shows results for three 569 node MIS problems encoded onto 20 qubits, all shown in Fig. \ref{fig:tested_graphs} g)-i).  Excluding infeasible solutions, ansatz B again performs significantly worse compared to the other three ansatze and compared to the previous tests. The ratios it produces do not exceed 0.74 and typical values are closer to 0.67.  Furthermore there is no improvement from increasing the number of repetitions.  Possible reasons for this plateauing in quality of the Heisenberg ansatz will be discussed in section \ref{sec:discussion}.  Surprisingly, ansatz C performs better here than for the case with 15 qubits.  While over 50\% of the solutions it generates are again infeasible, those that are feasible are on par with the solutions for ansatze A and D. For ansatz C we see another instance where increasing the number of repetitions can worsen the solution. Both the approximation ratio and the number of feasible solutions worsen going from two to three repetitions of ansatz C.  For ansatze A and D there is a very similar story as with the 15 qubit case.  The best results from both are the same across the three graphs(0.88, 0.89 and 0.85) but ansatz D improves in the consistency, producing less low-quality results than ansatz A.  Neither ansatz sees significant change from increasing the number of repetitions.
\subsection{Compression of information in the PCE}\label{sec:group_closure}
\begin{figure*}
    \centering
    \includegraphics[width=\linewidth]{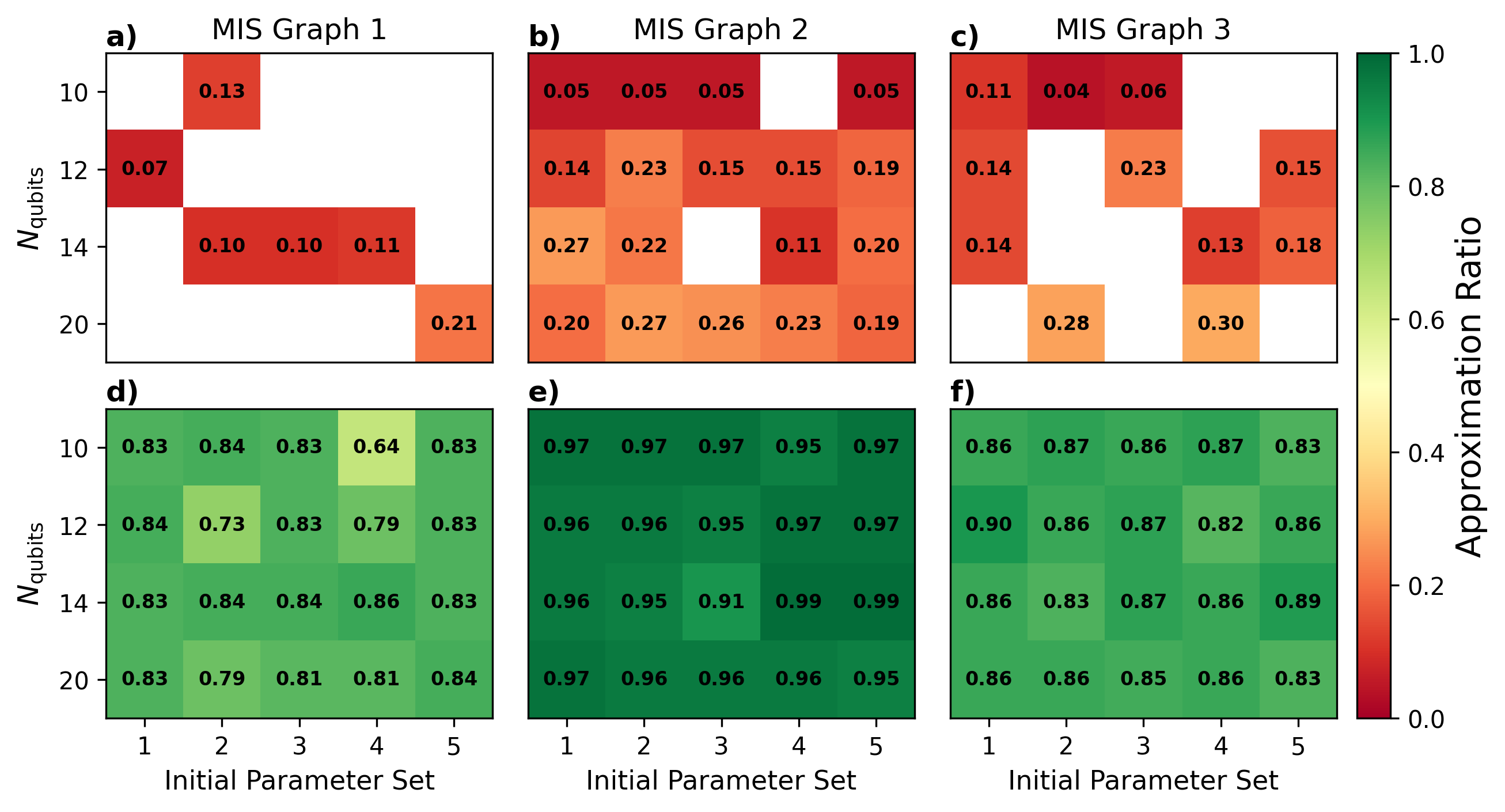}
    \caption{Approximation ratio produced by PCE before [a)-c)]and after [d)-f)] post processing using ansatz D in the three 134 node MIS graphs studied here[corresponding to results in Fig. \ref{fig:10_qubits} a)-d)] with five initial parameter seeds for each optimization.  y-axis indicates the number of qubits employed when creating the Pauli correlation encoding variables, 10, 12, 14 and 20 qubits are tested.  }
    \label{fig:group_closure}
\end{figure*}

In this section we look how the solution quality is affected by changing the portion of Pauli strings used in the PCE.  This is intended to explore how the quality of solutions is effected by the compression of information in the PCE.  We do this by increasing the number of qubits beyond the minimum needed for a given problem size and selecting a subset of the available strings.  Fig. \ref{fig:group_closure} shows the approximation ratio for the raw (a-c) and post-processed (d-f) values across the same three 134 node MIS instances studied in Fig. \ref{fig:10_qubits}.  We use ansatz D and employ PCE using 10, 12,14 and 20 qubits which corresponds to using roughly 100\%, 75\%, 50\% and 25\% of the total available Pauli strings respectively.

From these results we see that changing the portion of the total Pauli strings used has the strongest effect on the raw solutions. Although the number of feasible solutions stays roughly the same for each number of qubits the ratios are improved with an increasing number of available strings.  From 10 to 20 qubits the approximation ratios improve by 0.08, 0.22 and 0.19 for each graph respectively.  Unfortunately despite this improvement the raw values are still quite poor and post processing is still required.  Interestingly there is little change in the post-processed results across the different qubit numbers.  This highlights an important fact that ``good'' PCE solutions may seem very far from optimal but can be locally close to high quality solutions, in the sense that a optimal or near optimal solution can be recovered with a quick local bit-flip search.  Possible reasons for the improvement of the raw values with increased number of available Pauli strings will be discussed in Section \ref{sec:discussion}

\section{Discussion and Conclusions}\label{sec:discussion}
We will first discuss the results in section \ref{sec:ansatz_testing}.  Clearly, making the right choice of ansatz can have a huge impact on the solution quality.  Our experiments show that ansatz A and D are the best performing.  Both produce feasible solutions consistently and the best solutions from each are on par with one another.  However, ansatz D performs more consistently with smaller deviation between repeated runs in a fixed graph, in particular when scaling up the problem.  We attribute the superior performance of D over ansatz A to the fact that the former includes direct, parameterized coupling between $XX$, $YY$ and $ZZ$ on neighbor sites which we suggest allows the circuit to more directly tune the Pauli correlators allowing the classical optimizer to find better solutions compared to ansatz A.    For ansatz B, two differences exist between it and ansatz D that can explain it's worse performance.  Firstly, it uses $Rz$ single qubit rotations instead of $Ry$ and secondly, each of its entangling gates has its own unique optimization parameter.  The latter could lead to poor performance the larger parameter space leads the optimizer to get stuck.  However, if this were the sole issue we would expect for 10 qubits that ansatz A with 3 more parameters would also suffer performance issues.  Thus, we conclude that the use of the $Ry$ single qubit rotation in ansatz D aides in performance. This is likely due to less trivial action of the gate on computational basis states compared to the $Rz$ rotation in ansatz B.  As the problem is scaled to more qubits the number of optimization variables likely compound on the poor performance of ansatz B and add to the degradation of ansatz A.  This may also explain why adding more repetitions rarely lead to noticeably better results, while it can in principle increase the expressivity having too many parameters can lead to poor optimizer performance.  The scaling of parameter count with respect to qubit number and number of repetitions can be found in Table \ref{tab:parameter_scaling} for each ansatz. For ansatz C, which performed inconsistently and had a tendency to produce invalid solutions, we believe the limited number of parameters is too restrictive and despite converging quickly leads to poor overall performance.  Overall we conclude that in the tested simulated MIS instances, ansatz D produced strong and relatively consistent
results, and its slower parameter growth suggests it may be more scalable than ansatz A or B in similar settings(see Table \ref{tab:parameter_scaling}).

Next we turn to the results in Section \ref{sec:group_closure}.  One of the major issues with the PCE is that the group structure of Pauli operators can lead to dependence of Pauli correlators on one another which can lead to a restricted solution space\cite{xu2024boundingthejointgnumerical}.  The Pauli group's multiplicative structure means that
\begin{equation*}
    (Z_i Z_j )(Z_j Z_k) = Z_iZ_k
\end{equation*}
and similarly for $X$ and $Y$.  This means the expectation values used to encode our problems binary variables are not fully independent and as the expectation values move closer to classical values $\pm 1$ the solution space will become restricted due to these dependencies.  For example when the expectation values are $\pm 1$(or sufficiently close) we have
\begin{equation}
    \begin{split}
        x_1 &\sim \textrm{sgn}(\langle Z_iZ_j\rangle) = 1\\
        x_2 &\sim \textrm{sgn}(\langle Z_jZ_k\rangle) = -1\\
        \implies x_3 &\sim \textrm{sgn}(\langle Z_iZ_k\rangle) = x_1x_2 = -1
    \end{split}
\end{equation}

Since expectation values do not follow the multiplicative structure of the Paulis when they are far from classical values of $\pm 1$ this can be remedied by keeping the expectation values small, which is the purpose of the regularization term is included in equation \eqref{eq:pce_cost_function}.  In section \ref{sec:group_closure} we investigated another route to relieve this issue by increasing the number of available Paulis in the hope this would relieve some of the interdependence, allowing the solver to more freely search the solution space.  This approach found limited success, providing better raw results but offering little improvement to the post processed solutions.  This improvement may come from lowering the dependence induced by the group closure. It may also be the case that the dependence cannot be removed without returning to one-hot encoding, this is because it only takes $N_q-1$ order two Pauli strings to generate the remaining order two strings through multiplication.  Thus as soon as one has $N_q-1$ order two Pauli strings the dependence between variables is present. The use of fewer Pauli strings may still have an effect on the magnitude that the expectation values need in order to be independent but it cannot eliminate the dependence all together.  A second possibility for why the PCE saw improvement when more correlators are available is that when we increase the number of Pauli strings available there will be more available that contain operators acting on nearby qubits(i.e. more strings like $\dots IX_iX_{i+1}I\dots$ vs $X_i II\dots I X_j$) and the ansatz may simply be performing better on these strings. Since interdependence of correlators marks a significant barrier for the application of the PCE, mitigation of this issue is ongoing work.

\FloatBarrier
\appendix
\section{Maximum Independent Set Graphs}\label{sec:graph_instances}

\begin{figure*}[h!]
    \centering
    \includegraphics[width=\linewidth]{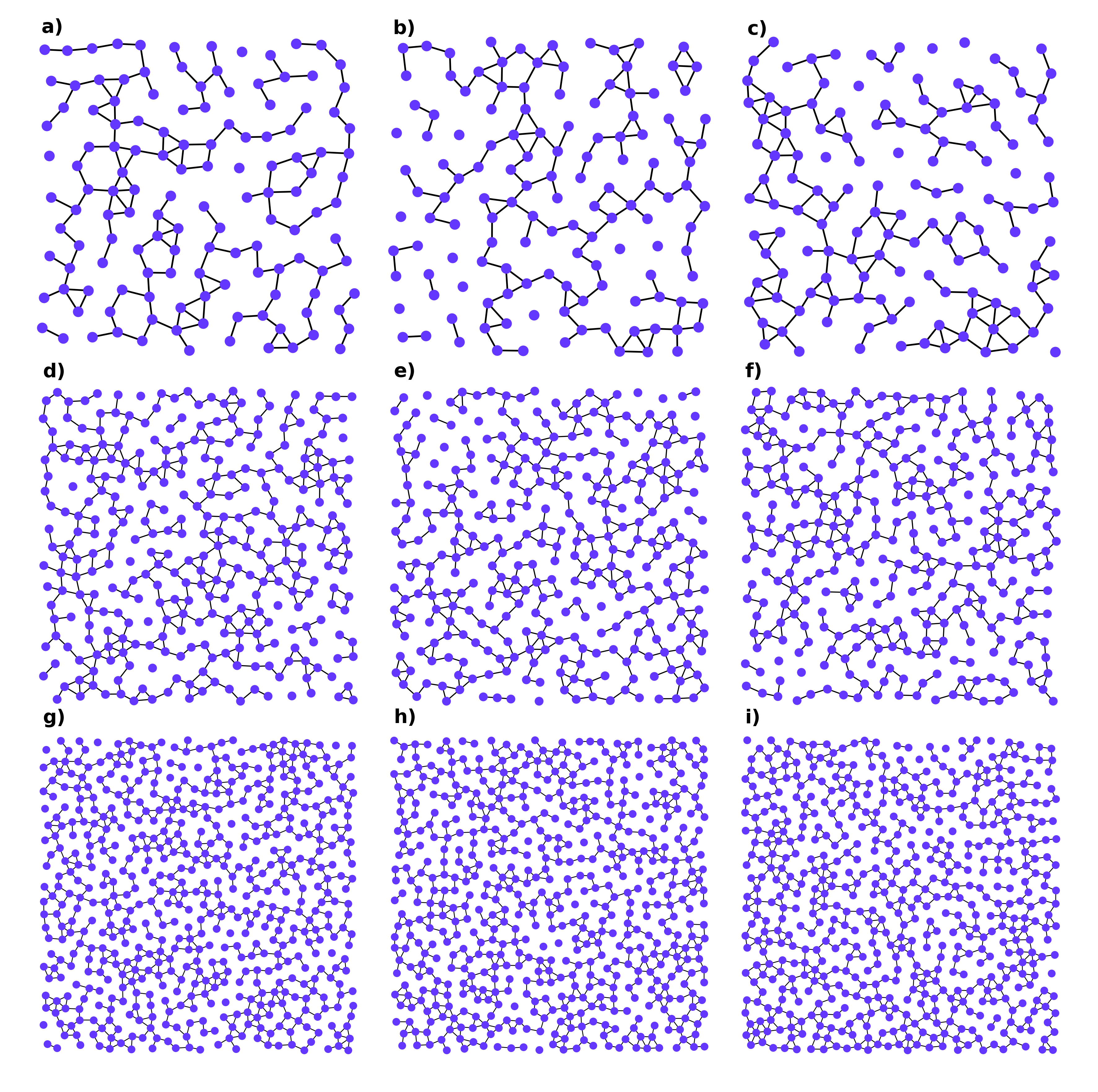}
    \caption{134(a-c), 314(d-f) and 569(g-i) node maximum independent set graphs tested }
    \label{fig:tested_graphs}
\end{figure*}
\clearpage 
\begin{center} 
{\large\bfseries References} 
\end{center} \vspace{0.5em} 
\bibliography{antenna_refs}

@misc{dinh2022efficienthamiltonianreductionquantum,
      title={Efficient Hamiltonian Reduction for Quantum Annealing on SatCom Beam Placement Problem}, 
      author={Thinh Q. Dinh and Son Hoang Dau and Eva Lagunas and Symeon Chatzinotas},
      year={2022},
      eprint={2210.06042},
      archivePrefix={arXiv},
      primaryClass={cs.IT},
      url={https://arxiv.org/abs/2210.06042}, 
}

@misc{krikidis2024optimizingconfigurationselectionreconfigurableantenna,
      title={Optimizing Configuration Selection in Reconfigurable-Antenna MIMO Systems: Physics-Inspired Heuristic Solvers}, 
      author={I. Krikidis and C. Psomas and A. K. Singh and K. Jamieson},
      year={2024},
      eprint={2406.17339},
      archivePrefix={arXiv},
      primaryClass={cs.IT},
      url={https://arxiv.org/abs/2406.17339}, 
}

@misc{macaluso2025quantumcomputinglargescalenetwork,
  title={Quantum Computing for Large-scale Network Optimization: Opportunities and Challenges},
  author={Sebastian Macaluso and Giovanni Geraci and Elías F. Combarro and Sergi Abadal and Ioannis Arapakis and Sofia Vallecorsa and Eduard Alarcón},
  year={2025},
  eprint={2509.07773},
  archivePrefix={arXiv},
  primaryClass={cs.NI},
  url={https://arxiv.org/abs/2509.07773}
}

@misc{vandelli2025constraintpreservingquantumalgorithmmultifrequency,
      title={Constraint-preserving quantum algorithm for the multi-frequency antenna placement problem}, 
      author={Matteo Vandelli and Francesco Ferrari and Daniele Dragoni},
      year={2025},
      eprint={2511.15566},
      archivePrefix={arXiv},
      primaryClass={quant-ph},
      url={https://arxiv.org/abs/2511.15566}, 
}

@misc{krikidis2025quantumoptimizationwirelesscommunication,
      title={Quantum Optimization in Wireless Communication Systems: Principles and Applications}, 
      author={Ioannis Krikidis and Valentin Gilbert},
      year={2025},
      eprint={2512.02468},
      archivePrefix={arXiv},
      primaryClass={cs.IT},
      url={https://arxiv.org/abs/2512.02468}, 
}

@misc{paul2026warmstartquantumapproximateoptimization,
      title={Warm-Start Quantum Approximate Optimization Algorithm for QAM MIMO Data Detection}, 
      author={Soumyadip Paul and Sourav Banerjee and Debanjan Bhowmik and Neel Kanth Kundu},
      year={2026},
      eprint={2604.18479},
      archivePrefix={arXiv},
      primaryClass={eess.SP},
      url={https://arxiv.org/abs/2604.18479}, 
}

@inproceedings{Kanza2025Spatially,
  title={Spatially Enhanced Quantum Optimization for Network Planning},
  author={Yaron Kanza and Divesh Srivastava},
  booktitle={Proceedings of The 33rd ACM International Conference on Advances in Geographic Information Systems (SIGSPATIAL '25)},
  year={2025},
  pages={5 pages},
  doi={10.1145/3748636.3760457}
}

@article{Lim2024QuantumAssisted,
  title={Quantum-Assisted Combinatorial Optimization for Reconfigurable Intelligent Surfaces in Smart Electromagnetic Environments},
  author={Qi Jian Lim and Charles Ross and Amitabha Ghosh and Frederick W. Vook and Gabriele Gradoni and Zhen Peng},
  journal={IEEE Transactions on Antennas and Propagation},
  year={2024},
  volume={72},
  number={1},
  pages={147--159},
  doi={10.1109/TAP.2023.3298134}
}

@inproceedings{Arani2025Quantum,
  title={Quantum Machine Learning for Optimal Antenna Placement},
  author={Masoud Salmani Arani and Reza Shahidi and Lihong Zhang},
  booktitle={2025 IEEE International Symposium on Antennas and Propagation and North American Radio Science Meeting (AP-S/CNC-USNC-URSI)},
  year={2025},
  pages={1499--1502},
  doi={10.1109/AP-S/CNC-USNC-URSI55537.2025.11266785}
}

@article{Bouchmal2025Quantum,
  title={Quantum computing approach for multi-objective routing and spectrum assignment optimization},
  author={Oumayma Bouchmal and Bruno Cimoli and Ripalta Stabile and Juan Jose Vegas Olmos and Carlos Hernandez-Chulde and Ricardo Martínez and Ramon Casellas and Idelfonso Tafur Monroy},
  journal={Journal of Optical Communications and Networking},
  year={2025},
  volume={17},
  number={6},
  pages={B15--B27},
  doi={10.1364/JOCN.552061}
}

@article{Takabayashi2025Optimization,
  title={Optimization of connection patterns between mobile phones and base stations using quantum annealing},
  author={T. Takabayashi and S. Sudo and T. Aoki and others},
  journal={Scientific Reports},
  year={2025},
  volume={15},
  pages={24540},
  doi={10.1038/s41598-025-09230-5}
}

@article{Batool2026QuantumEnhanced,
  title={Quantum-Enhanced Massive MIMO Beamforming for 6G IoT Networks: A QAOA-Based Optimization Framework},
  author={Iqra Batool and Mostafa M. Fouda and Muhammad Ismail and Mohamed I. Ibrahem and Zubair Md Fadlullah and Nei Kato},
  journal={IEEE Open Journal of the Communications Society},
  year={2026},
  volume={7},
  pages={222--238},
  doi={10.1109/OJCOMS.2025.3645207}
}

@article{Original_Sciorilli_2025,
   title={Towards large-scale quantum optimization solvers with few qubits},
   volume={16},
   ISSN={2041-1723},
   url={http://dx.doi.org/10.1038/s41467-024-55346-z},
   DOI={10.1038/s41467-024-55346-z},
   number={1},
   journal={Nature Communications},
   publisher={Springer Science and Business Media LLC},
   author={Sciorilli, Marco and Borges, Lucas and Patti, Taylor L. and García-Martín, Diego and Camilo, Giancarlo and Anandkumar, Anima and Aolita, Leandro},
   year={2025},
   month=Jan }

@misc{carmo2025warmstartingpcetravelingsalesman,
      title={Warm-Starting PCE for Traveling Salesman Problem}, 
      author={Rafael S. do Carmo and Renato Gomes dos Reis and Samuel Fernando F Silva and Luiz Gustavo E. Arruda and Felipe F. Fanchini},
      year={2025},
      eprint={2509.14414},
      archivePrefix={arXiv},
      primaryClass={quant-ph},
      url={https://arxiv.org/abs/2509.14414}, 
}

@article{Soloviev2026LargeScale,
  title={Large-scale portfolio optimization using Pauli correlation encoding},
  author={V. P. Soloviev and M. Krompiec},
  journal={Scientific Reports},
  year={2026},
  doi={10.1038/s41598-026-54244-2}
}

@article{PadinMartinez2026Progressive,
  title={Progressive Binarization - Pauli Correlation Encoding: a Continuation Method for Constrained Optimization},
  author={Jacobo Padín-Martínez and Vicente P. Soloviev and Alejandro Borrallo-Rentero and Antón Rodríguez-Otero and Raquel Alfonso-Rodríguez and Michal Krompiec},
  journal={arXiv preprint arXiv:2602.17479},
  year={2026}
}

@misc{soloviev2026quantumlearningclassicalcorrelations,
      title={Quantum Learning of Classical Correlations with continuous-domain Pauli Correlation Encoding}, 
      author={Vicente P. Soloviev and Bibhas Adhikari},
      year={2026},
      eprint={2604.05637},
      archivePrefix={arXiv},
      primaryClass={quant-ph},
      url={https://arxiv.org/abs/2604.05637}, 
}

@misc{nguyen2026scalingquantumoptimizationunit,
      title={Scaling Quantum Optimization for Unit Commitment via Pauli Correlation Encoding}, 
      author={Kien X. Nguyen and Ilya Safro and Xiaoyuan Liu},
      year={2026},
      eprint={2605.17145},
      archivePrefix={arXiv},
      primaryClass={quant-ph},
      url={https://arxiv.org/abs/2605.17145}, 
}

@misc{Galda_mRNA_2026paulicorrelationencodingmrna,
      title={Pauli Correlation Encoding for mRNA Secondary Structure Prediction: Problem-Aware Decoding for Dense-Constraint QUBOs}, 
      author={Triet Friedhoff and Mihir Metkar and Wade Davis and Vaibhaw Kumar and Alexey Galda},
      year={2026},
      eprint={2605.20163},
      archivePrefix={arXiv},
      primaryClass={quant-ph},
      url={https://arxiv.org/abs/2605.20163}, 
}

@misc{alonso2026benchmarkpaulicorrelationencoding,
      title={Benchmark of Pauli Correlation Encoding for different optimisation problems}, 
      author={Fernando Alonso and Colomán Samprón and Jacobo Veiga and Mariamo Mussa Juane and Andrés Gómez},
      year={2026},
      eprint={2606.18914},
      archivePrefix={arXiv},
      primaryClass={quant-ph},
      url={https://arxiv.org/abs/2606.18914}, 
}

@misc{bosco2026efficientlysimulablepaulicorrelation,
      title={Efficiently Simulable Pauli Correlation Encoding}, 
      author={Daniele Lizzio Bosco and Gabriel Matos and Chen-Yu Liu and Frederic Rapp and Fabian Finger and Enrico Rinaldi and Konstantinos Meichanetzidis},
      year={2026},
      eprint={2607.20409},
      archivePrefix={arXiv},
      primaryClass={quant-ph},
      url={https://arxiv.org/abs/2607.20409}, 
}

@article{bharti2022noisyintermediate,
  title = {Noisy intermediate-scale quantum algorithms},
  author = {Bharti, Kishor and Cervera-Lierta, Alba and Kyaw, Thi Ha and Haug, Tobias and Alperin-Lea, Sumner and Anand, Abhinav and Degroote, Matthias and Heimonen, Hermanni and Kottmann, Jakob S. and Menke, Tim and Mok, Wai-Keong and Sim, Sukin and Kwek, Leong-Chuan and Aspuru-Guzik, Al\'an},
  journal = {Rev. Mod. Phys.},
  volume = {94},
  issue = {1},
  pages = {015004},
  numpages = {69},
  year = {2022},
  month = {Feb},
  publisher = {American Physical Society},
  doi = {10.1103/RevModPhys.94.015004},
  url = {https://link.aps.org/doi/10.1103/RevModPhys.94.015004}
}

@Article{Chen2023thecomplexityofnisq,
author={Chen, Sitan
and Cotler, Jordan
and Huang, Hsin-Yuan
and Li, Jerry},
title={The complexity of NISQ},
journal={Nature Communications},
year={2023},
month={Sep},
day={26},
volume={14},
number={1},
pages={6001},
issn={2041-1723},
doi={10.1038/s41467-023-41217-6},
url={https://doi.org/10.1038/s41467-023-41217-6}
}

@article{xu2024boundingthejointgnumerical,
  title = {Bounding the Joint Numerical Range of Pauli Strings by Graph Parameters},
  author = {Xu, Zhen-Peng and Schwonnek, Ren\'e and Winter, Andreas},
  journal = {PRX Quantum},
  volume = {5},
  issue = {2},
  pages = {020318},
  numpages = {19},
  year = {2024},
  month = {Apr},
  publisher = {American Physical Society},
  doi = {10.1103/PRXQuantum.5.020318},
  url = {https://link.aps.org/doi/10.1103/PRXQuantum.5.020318}
}

\end{document}